\documentclass[aps,prl,twocolumn,groupedaddress,showpacs,floatfix]{revtex4-1}

\usepackage{amsmath}
\usepackage{graphicx}
\usepackage[usenames,dvipsnames]{color}
\usepackage{tabularx}
\graphicspath{{figs/}{:Figs:}} 
\begin{document}
\title{Exchange parameters and adiabatic magnon energies from spin-spiral calculations}

\author{Adam Jacobsson$^{1,2}$}

\author{Biplab Sanyal$^{2}$}

\author{Marjana Le\v{z}ai\'c$^{1}$}

\author{Stefan Bl\"ugel$^{1}$}

\affiliation{$^{1}$Peter Gr\"unberg Institute \& Institute for Advanced Simulation, Forschungzentrum J\"ulich and JARA, 52425 J\"ulich, Germany}

\affiliation{$^{2}$Department of Physics and Astronomy, Uppsala University, Box-516,75120, Uppsala, Sweden}

\date{\today}

\begin{abstract}
We present a method of extracting the exchange parameters of the classical Heisenberg model from first-principles calculations of spin-spiral total energies based on density functional theory. The exchange parameters of the transition-metal monoxides MnO and NiO are calculated and used to estimate magnetic properties such as transition temperatures and magnon energies. Furthermore we show how to relate the magnon energies directly to differences in spin-spiral total energies for systems containing an arbitrary number of magnetic sublattices. This provides a comparison between magnon energies using a finite number of exchange parameters and the infinite limit.

\end{abstract}

\maketitle

\section{Introduction}

A crucial task in the field of theoretical magnetism is the prediction of non-zero temperature properties and especially magnetic transition temperatures. A fruitful solution to these problems has been proposed by assuming that the magnetic excitations can be described  by a Heisenberg Hamiltonian with exchange parameters obtained from density functional theory (DFT). The procedure rests on the adiabatic assumption that the time scale of magnon- and electronic motion differs enough to let the local electronic structure adapt to the presence of magnons.  This assumption allows one to deal with magnons as frozen in spin-spiral modulations of the ground state magnetic structure and to calculate the total energies within constrained non-collinear DFT~\cite{Halilov1998}.  While adiabatic magnon dispersion curves have been calculated with the use of the frozen magnon technique for some time we contribute to the method by showing how to relate magnon energies directly to spin-spiral total energy differences for systems containing multiple magnetic sublattices. 

We describe our recent implementation of the least square fitting (LSF) approach for calculating the real space exchange parameters from the spin-spiral total energies and compare it to the approach where these parameters are obtained from their Fourier transforms (FT) in the reciprocal space~\cite{Lezaic2012}. Both approaches were implemented in the full-potential linearized augmented planewave (FLAPW) method-based code {\tt FLEUR}~\cite{FLEUR}.The least square fitting scheme is investigated in some detail and is shown to reduce the computational costs in some cases while obtaining identical results as the Fourier transformation based approach. The methods are applied to the the transition-metal monoxides, viz., NiO and MnO.  

NiO and MnO adopt the rocksalt structure in the paramagnetic phase. Below the N\`eel temperature an antiferromagnetic magnetic ordering sets in where the direction of the atomic moments alternates between neighboring [111] planes. Exchange-striction leads to a simultaneus structural phase transition where the rocksalt structure is distorted into a trigonal one. 

Both MnO and NiO are well known for having strong correlation effects associated with the 3$d$-electrons localized on the transition-metal ions. DFT functionals such as the local density approximation (LDA) and the generalized gradient approximation (GGA) are not able to describe strong electron correlation. Beyond DFT methods such as the LDA+U~\cite{Anisimov:1991ys,Solovyev:1998fv, Bengone:2000zr, Rohrbach:2004ly, Zhang:2006ve, Karlsson:2010kx}, Self interaction correction (SIC)~\cite{Szotek:1993bh, Svane:1990qf, Kodderitzsch:2002dq}, dynamical mean field theory (DMFT)~\cite{Kunes:2007oq}, hybrid funtionals~\cite{Franchini:2005cr, Feng:2004nx,Schlipf:2011uq} and GW-approximation \cite{Faleev:2004vn, Kotani:2008uq} have been employed with greater success improving the correspondence between calculated and experimental properties such as lattice parameters, band gaps and excitation energies for magnons and phonons. In this study we employ the LDA+U method for this class of materials.

As an application of our method we show how the selection of the spin spirals can be made to reduce the computational cost compared to the previously implemented FT based method.~\cite{Lezaic2012} The gains achieved are expected to be larger in the case of insulators where the number of interactions are relatively small. 

Furthermore we show how the magnon dispersion curves depend on the number of exchange parameters used in the least square approach and compare it to the curves obtained directly from spin-spiral total energy calculations. In addition, we calculate the magnetic transition temperatures using Monte Carlo simulations for the transition-metal monoxides. It is shown that the calculated exchange parameters and consequently the transition temperatures and magnon dispersion curves  depend significantly on the crystal structure (i.e., the ideal rock salt or the trigonal one) for MnO. It is also shown that the LDA+U functional in the full potential implementation gives a good description of the magnetic properties of the transition-metal monoxides NiO and MnO for a particular set of values of Hubbard U in contrast to former atomic-sphere-approximation (ASA) results where no such values could be found.~\cite{Solovyev:1998fv}

\section{Theory}
\subsection{Exchange parameters}
We employ a classical Heisenberg model where normalized vector spin moments ${\bf e}_{n\alpha}$ are localized at ionic sites  ${\bf R}_{n\alpha}$ defined by a lattice vector ${\bf R}_{n}$ and position vector ${\bf \tau}_{\alpha}$ of the magnetic Bravais lattice within a unit cell.
\begin{equation}
{\bf R}_{n\alpha} = {\bf R}_{n} +{\bf \tau}_{\alpha}
\end{equation}
\\
The spins interact via exchange coupling parameters $J_{mn}^{\alpha\beta}$	and the exchange Hamiltonian $H_{ex}$ is the sum over all pair interactions.
\begin{equation}
H_{ex}=-\frac{1}{2N}\sum_{mn\alpha \beta} J^{\alpha \beta}_{mn} { \bf e}_{m\alpha}\cdot {\bf e}_{n\beta}
\label{eqn2}
\end{equation}

Here $N$ is the number of unit cells in the crystal. Possible ground states of a Hamiltonian of the form ~\eqref{eqn2} are spin spirals defined by wave vectors {\bf q} from the irreducible wedge of the Brillouin zone  together with angles $\theta$ between the spin and rotation axis along with  a phase factor  $\phi$ common for all atoms belonging to the same magnetic Bravais lattice~\cite{bertaut:1138}. For single ${\bf q}$-states the spin ${\bf e}_{m\alpha}$ has the form:

\begin{equation}
\begin{split}
{\bf e}_{m\alpha} & = \sin(\theta) \cos(\gamma_{m\alpha})\mathrm{\bf x} + \sin(\theta) \sin(\gamma_{m\alpha})\mathrm{\bf y} + \cos(\theta)\mathrm{\bf z} \\
\gamma_{m\alpha} & ={\bf q}\cdot {\bf R}_{m\alpha} +\phi_{\alpha}
\end{split}
\label{eqn3}
\end{equation}

A commonly used method to extract the exchange parameters from DFT is to assume a maximum range of interactions and solve the system of equations given by the Hamiltonian ~\eqref{eqn2} using total energies E({\bf q}) of different collinear magnetic configurations, i.e.\ plane spin spirals for high symmetry {\bf q}-points.

\begin{equation}
E({\bf q})=-\frac{1}{2N}\sum_{mn\alpha \beta} J_{mn}^{\alpha \beta}{\bf e}_{m\alpha} ({\bf q}_{hs}) \cdot {\bf e} _{n\beta}({\bf q}_{hs})
\label{eqn4}
\end{equation}

A problem with this approach is that deviations from the Heisenberg model can be expected for such large perturbations from the ground state.~\cite{Halilov1998} This can for instance be seen in the changing magnitude of the atomic spin moments for different collinear configurations. It may thus be advantageous to extract the exchange parameters from cone spin spirals which exert smaller perturbations. From Eq. ~\eqref{eqn3} and \eqref{eqn4} one obtains the relation between the {\em ab initio} total energies and the exchange parameters.

\begin{equation}
\begin{split}
& E[{\bf q},(\theta_{1} ,...,\theta_{l} ),(\phi_{1} ,...,\phi_{l})] = E_{0} - \frac{1}{2}\sum_{n\alpha\beta} J^{\alpha \beta}_{0n}\times\\
& (\sin(\theta_{\alpha})\sin(\theta_{\beta})\cos({\bf q} \cdot ({\bf R}_{0\alpha}- {\bf R}_{n\beta}) + \phi_{\alpha}-\phi_{\beta})\\
&-\cos(\theta_{\alpha})\cos(\theta_{\beta}))
\end{split}
\label{eqn5}
\end{equation}

We will make use of the following notation: 

\begin{equation}
 E_{\alpha\beta}^{\phi}[{\bf q},\{ \theta\}] = E[{\bf q},(\theta_{1} ,...,\theta_{l} ),(\phi_{1} ,...,\phi_{l})]
\end{equation}
with the assumptions that the only non-zero cone angles are those of atoms belonging to the sublattices $\alpha$ and $\beta$ with $\theta_{\alpha}=\theta_{\beta}=\theta$. Furthermore the only non-zero phase factor is assigned to the magnetic moment of the sublattice $\alpha$, so that $\phi_{\alpha}=\phi$.  We use the following spin-spiral total energies for the fitting procedure. 

\begin{equation}
\begin{split}
&\frac{2}{sin^2(\theta)}(E^0_{\alpha\alpha}[{\bf 0},\{ \theta\}]- E^0_{\alpha\alpha}[{\bf q},\{ \theta\}])\\
& = \sum_{n} J^{\alpha \alpha}_{0n} ~ (1-\cos({\bf q}\cdot{\bf R}_{n\alpha}))
\end{split}
\label{eqn6}
\end{equation}

\begin{equation}
\begin{split}
&\frac{2}{sin^2(\theta)}(E^0_{\alpha\beta}[{\bf 0},\{ \theta\}]- E^0_{\alpha\beta}[{\bf q},\{ \theta\}]) \\
& =  2\sum_{n} J^{\alpha\beta}_{0n} ~ (1-\cos({\bf q}\cdot({\bf R}_{0\alpha} - {\bf R}_{n\beta})))\\
&+ \sum_{n} J^{\alpha \alpha}_{0n} (1-\cos({\bf q}\cdot{\bf R}_{n\alpha}))\\
&+ \sum_{n} J^{\beta \beta}_{0n} (1-\cos({\bf q}\cdot{\bf R}_{n\beta}))\\
\end{split}
\label{eqn7}
\end{equation}

For each ${\bf q}$-point we obtain Eq.  ~\eqref{eqn6} for each magnetic sublattice and Eq.  ~\eqref{eqn7}  for each pair of magnetic sublattices. In principle the latter kind is enough to obtain all exchange parameters by solving the system of equations but we use both equations in order to facilitate comparisons with the FT based approach and to reduce the number of ${\bf q}$-points. 

In practice we calculate spin-spiral total energies for each ${\bf q}$ where we put a single non-zero cone angle $\theta$ on each of the sublattices in turn. In addition we calculate the spin-spiral total energies for each ${\bf q}$ where we put a cone angle $\theta$ on both atoms for each pair of sublattices. 

In order to improve the quality of the calculations and facilitate the human effort of executing them, a least square fitting procedure is applied to obtain the parameters, given that an equal or bigger number of spin-spiral total energies is supplied. We use the singular value decomposition (SVD) technique~\cite{Atkinson1989} since it is known to be robust even for least square fitting problems that are close to be rank- deficient, i.e.\ where the number of linearly independent rows is less than the number of columns.

The condition number is a measure of the closeness to rank deficiency and is given from the SVD as the quotient of the largest and smallest singular value of the matrix of the problem. For a rank deficient matrix, the smallest singular value goes to zero whereas the condition number goes to infinity. With a high condition number the fitted parameters are sensitive to perturbation in the input data, which in our case is the difference in total energy between spin spirals. That means that the evaluation of the exchange parameters puts different demands on the precision of the DFT calculations for different sets of spin spirals. The conditioning of the numerical problem is a property of the selected set of ${\bf q}$-points and can be established prior to any actual ab-initio calculation. In this way we can ensure that meaningful results are obtained without spending any significant amount of computing time. 

Compared to the FT based method, the choice of the sampling of the Brillouin zone is more flexible for the LSF, since the discrete Fourier transform requires the total energies of a set of {\it equidistant} wave vectors that sample the whole Brillouin zone while in the present case, we can solve the system of equations for {\it any set} of wave vectors as long as the matrix representation of the problem is not rank deficient. \\

\subsection{Magnons}
From classical spin-wave theory, one can derive magnon frequencies $\omega_{\bf q}$. from the information provided by the exchange parameters. We obtain $\omega_{\bf q}$ for collinear spin-configurations as the positive eigenvalues of the spin-wave dynamical matrix $\Delta({\bf q})$ ~\cite{Halilov1998}.

\begin{equation}
\Delta_{\alpha \beta}({\bf q}) = 2\left(\delta_{\alpha \beta}\sum_{\gamma}\frac{J^{\alpha \gamma}({\bf 0})M_{\gamma}}{|M_{\alpha}||M_{\gamma}|}- \frac{J^{\alpha \beta}({\bf q})M_{\beta}}{|M_{\alpha}||M_{\beta}|}\right)
\label{eqn8}
\end{equation}

\begin{equation}
J^{\alpha \beta}({\bf q})=\sum_{n}J^{\alpha \beta}_{0n}\cos({\bf q} \cdot ({\bf R}_{0\alpha}- {\bf R}_{n\beta})) 
\end{equation}

In the case of  a single magnetic sublattice, i.e. a simple ferromagnet, this expression reduces to the following familiar form with quadratic dispersion close to $\Gamma$.

\begin{equation}
\omega_{\bf q} = 2\dfrac{J({\bf 0})- J({\bf q})}{M}
\label{1magnon}
\end{equation}

Since the Fourier transformed exchange constants are directly related to spin-spiral total energies,~\cite{Lezaic2012} it is possible to calculate magnon energies without explicit calculations of the real space exchange parameters. The spin-wave dynamical matrix is formed with matrix elements given directly from energy differences of spin-spiral total energies obtained by a straight forward comparison of Eq.  \eqref{eqn5} and \eqref{eqn8}.  The diagonal and off-diagonal elements are respectively given by the following equations:

\begin{equation}
\begin{split}
&\Delta_{\alpha \alpha}({\bf q}) = \frac{2}{|M_{\alpha}|\sin^2(\theta)}\times \bigg(\frac{2M_{\alpha}}{|M_{\alpha}|}(E_{\alpha\alpha}^{0}[{\bf q},\{\theta\}  ]\\
& -E_{\alpha\alpha}^{0}[{\bf 0},\{\theta\}  ] ) +\sum_{\gamma \neq \alpha}\frac{M_{\gamma}}{|M_{\gamma}|}(E_{\alpha\gamma}^{\pi /2}[{\bf 0},\{\theta\}  ]-E_{\alpha\gamma}^{0}[{\bf 0},\{\theta\}  ])\bigg)
\end{split}
\label{emagnon1}
\end{equation}

\begin{equation}
\begin{split}
& \Delta_{\alpha \beta}({\bf q}) = \frac{2}{|M_{\alpha}|\sin^2(\theta)}\times \bigg(\frac{M_{\beta}}{|M_{\beta}|}(E_{\alpha\beta}^{0}[{\bf q},\{\theta\}  ]\\
& -E_{\alpha\beta}^{\pi /2}[{\bf 0},\{\theta\}  ])- \sum_{\gamma=\alpha,\beta} \frac{M_{\gamma}}{|M_{\gamma}|}(E_{\gamma\gamma}^{0}[{\bf q},\{\theta\}  ]-E_{\gamma\gamma}^{0}[{\bf 0},\{\theta\}  ])\bigg)
 \end{split}
\label{emagnon2}
\end{equation}


In the case of a single magnetic sublattice, Eq.  \eqref{emagnon1} reduces to the following well known expression:

\begin{equation}
\omega_{\bf q} = 4 \frac{E({\bf q},\theta)- E({\bf 0},\theta)}{M \sin^2(\theta)}
\label{11magnon}
\end{equation}

We note that our expressions  \eqref{emagnon1}  and  \eqref{emagnon2}  deviates from those in a recent publication.~\cite{Essenberger:2011kx}

\section{Computational details}
We use experimental lattice parameters in all calculations except that we neglect the small trigonal distortion whenever it's not mentioned explicitly. The lattice parameters considered in our calculations are 4.16 \AA \ and 4.44 \AA \ for NiO and MnO respectively. 

\textit{Ab initio} total energies were obtained by the FLAPW code  {\tt FLEUR}~\cite{FLEUR}. We used the Perdew-Zunger LDA~\cite{PhysRevB.23.5048}  exchange-correlation functional to which we added a Hubbard U according to the formulation in Ref.~\cite{Shick:1999vn}. The double counting correction was taken to satisfy the fully localized limit.~\cite{Kugel:1978kl} Several different methods of extracting Hubbard U and Hund's J parameters from ab initio calculations have been developed such as constrained LDA calculations~\cite{Anisimov:1991ys}, linear response calculations~\cite{Cococcioni:2005fk} and the constrained random phase approximation~\cite{Karlsson:2010kx}. NiO is a common benchmark material and the values of U and J have been calculated for all the above mentioned methods. The values of U range from 4.6 eV to 8.0 eV depending on the details of the method of calculation and definition of the localised orbitals that are treated with the Hubbard U. Results for the calculations of Hubbard U and Hund's J for MnO are less common in the literature where we only found the results of constrained LDA calculations. 

There has been an extensive discussion of what magnitude of Hubbard U produces the best correspondence with different experimental properties of NiO: it was argued that a better estimation of structural parameters, correspondence with electron-energy-loss spectra and optical propertied were obtained with a Hubbard U in the range 5-6 eV rather than the 8 eV obtained in constrained LDA-calculations~\cite{Dudarev:1998ys, Bengone:2000zr}.  

In this study we did not calculate the Hubbard U but used the values taken from previous constrained LSDA+U calculations by Anisimov {\it et al.}\cite{Anisimov:1991ys}. These calculations gave U and J equal to 8.0 eV and 0.95  eV respectively for NiO and U and J equal to 6.9 eV and 0.86 eV respectively for MnO. These choice of parameters fixed the functional for the calculations of exchange parameters and derived properties such as magnon dispersion and critical temperatures. However we completed the study by also calculated the exchange parameters considering a scan over the U-values from 3 to 9 eV for these two oxides. 

The non-collinear magnetism in ${\tt FLEUR}$ is implemented in the atomic moment approximation, which assumes an intra-atomic collinearity~\cite{Kurz:2004fk}. A grid of  25$\times$25$\times$25 k-point mesh~\cite{Monkhorst:1976fk} was considered. We used a muffin-tin radius of 1.19 \AA. for the transition-metal atoms and 0.83~\AA\ for the oxygen. The planewave cutoff was fixed by setting the $k_{max}$ parameter to 8.38 \AA$^{-1}$. A cone angle $\theta$ of $\pi / 6$ was used throughout the work. The calculations of exchange constants were converged to a precision of 0.1 meV with respect to the parameters considered above for both materials. 

\begin{figure}[h]
\begin{center}
\includegraphics[width=1.0\columnwidth]{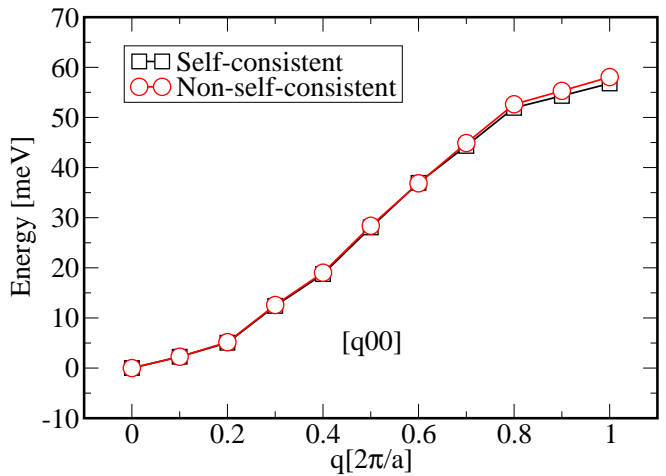}
\caption{(Color online) A comparison between spin-spiral total energies calculated with the LDA+U functional self consistently and non-self consistently using the force theorem for NiO. The parameters U and J were set to 8.0 eV and 0.95 eV respectively.}
\label{NiOdisp}
\end{center}
\end{figure}

To reduce computational expense, non-self consistent calculations of the spin-spirals were employed and the total energy differences between two spin spirals  were approximated by the differences of the sum of eigenvalues as described by Andersen's force theorem~\cite{Liechtenstein:1987fk}. The reliability of the non-self consistent total energy calculations was checked and is shown in Fig.~\ref{NiOdisp} for the case of NiO.  It is clearly observed that only relatively small deviations are found at the Brillouin zone boundary.

\section{Results}

\begin{figure}[h]
\begin{center}
\includegraphics[width=0.35\textwidth]{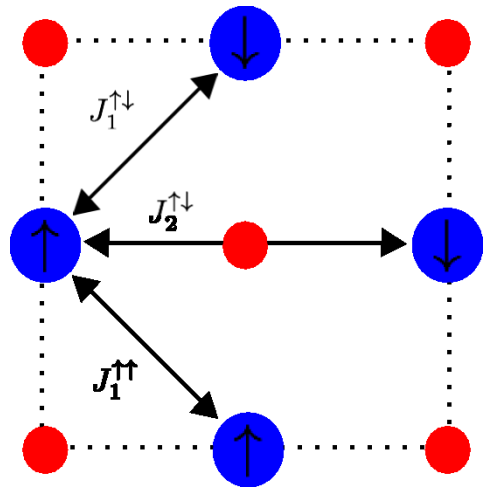}
\caption{(Color online)The nearest and next nearest exchange interactions. Blue balls represent transition metal ions and red balls represent oxygen ions.}
\label{Rocksalt}
\end{center}
\end{figure}

\begin{figure}[h]
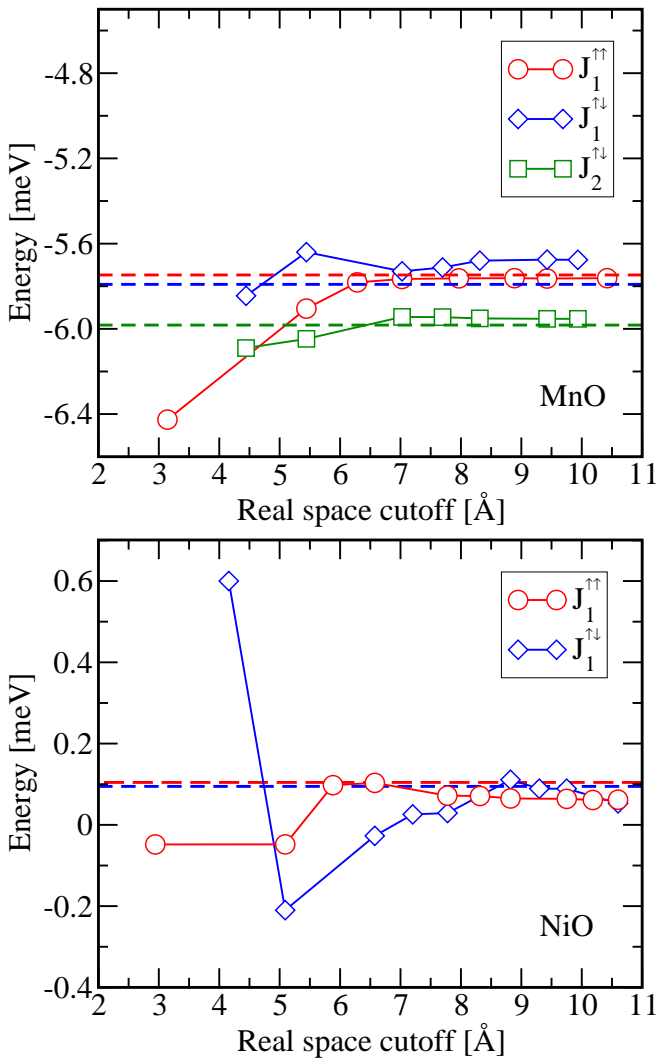

\centering
\begin{tabular}{c}
\includegraphics*[width=1.0\columnwidth]{MnOC.eps}\\
\includegraphics*[width=1.0\columnwidth]{NiOC.eps}
\end{tabular}
\caption{(Color online) The convergence of the exchange parameters with respect to the real space cutoff. We include the corresponding values for the exchange parameters calculated with the Fourier transform by dashed lines. The parameters U and J were set to respectively 6.9 eV and 0.86 eV for MnO and 8.0 eV and 0.95 eV for NiO.}
\label{TmOC}
\end{figure}

\begin{figure}[h]
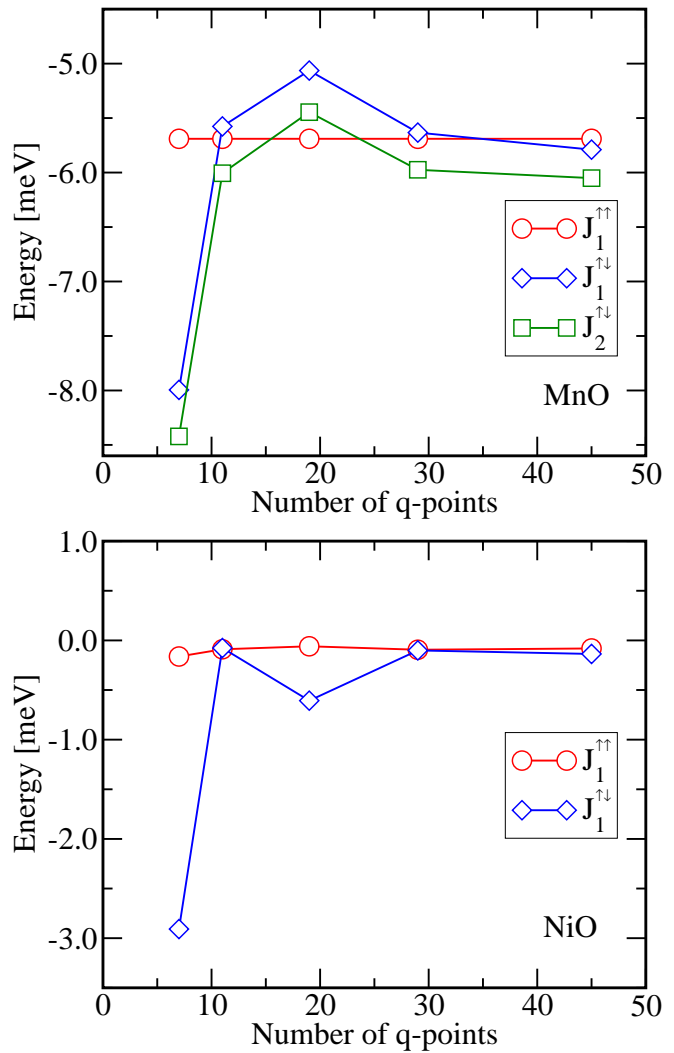

\centering
\begin{tabular}{c}
\includegraphics*[width=1.0\columnwidth]{MnOFC.eps}\\
\includegraphics*[width=1.0\columnwidth]{NiOFC.eps}
\end{tabular}
\caption{(Color online) The convergence of the exchange parameters with number of q-points using ${\bf q}$-points using the FT based method. The parameters U and J were set to respectively 6.9 eV and 0.86 eV for MnO and 8.0 eV and 0.95 eV for NiO.}
\label{TmOFC}
\end{figure}

The unit cell of the antiferromagnetic configurations contains two magnetic sublattices. The exchange parameters for atomic moments aligned in the same and opposite directions were extracted with the LSF and FT. In Fig.~\ref{TmOC} we show the convergence of the nearest and second nearest neighbor exchange parameter for MnO while only the nearest neighbor is shown for NiO. The notation of Fig.~\ref{TmOC} is clarified in Fig.~\ref{Rocksalt}. 

In order to directly compare the LSF and the previously implemented FT based method~\cite{Lezaic2012} we choose to calculate the total energies of ${\bf q}$-points distributed in an equidistant mesh~\cite{Monkhorst:1976fk} in the irreducible Brillouin zone. To reach the desired accuracy of 0.1 meV we had to calculate 11 ${\bf q}$-points for the LSF for both materials. However as we did our convergence tests we saw that the results using the FT were not stable with respect to increases in the number ${\bf q}$-points until we reached the set of 29 ${\bf q}$-points. In Fig.~\ref{TmOFC} we show the convergence of the FT-based method. 

It can be seen in Fig.~\ref{TmOC} that the results obtained by the LSF converge within an energy interval of 0.1 meV only with the inclusion of exchange parameters beyond the second nearest neighbor. The transition-metal monoxides have exchange parameters of a magnitude larger than 0.1 meV beyond the second nearest neighbor which will be projected on and change the nearest and second nearest neighbor if neglected. This finding is different from recently reported results obtained with the LDA-SIC functional \cite{Fischer:2009bs}, where no interaction beyond the second nearest neighbor was significant. 

Table I.  contains the calculated exchange parameters with a magnitude larger than 0.1 meV using a full equidistant grid of ${\bf q}$-points. Here we also include results for $J_1$ and $J_2$ as calculated from the total energies of collinear states. We see that while the nearest and second nearest neighbor interaction parameters are close to the ones obtained by spin-spiral calculations in the case of NiO, it is not so for MnO where the extraction of the parameters from collinear calculations introduces errors of the order of meV. The exchange parameters should conform to he symmetry of the crystal.~\cite{Lezaic2012}. This means that $J^{\uparrow\uparrow}_1$ and $J^{\uparrow\downarrow}_1$ should be equal by symmetry in the ideal rock salt structure. The difference of the two curves $J^{\uparrow\uparrow}_1$ and $J^{\uparrow\downarrow}_1$ can thus be used as an estimation of the quality of the calculations.

\begin{table}[htb]
\caption{The exchange parameters larger than 0.1 meV for NiO and MnO calculated with the ideal rock salt structure. All exchange parameters are given in meV. The parameters U and J were set to respectively 6.9 eV and 0.86 eV for MnO and 8.0 eV and 0.95 eV for NiO.}
\begin{center}
\begin{tabular}{ccccc}
\hline
  \hline
  & & & \\
 Spin-Spiral Calculations & MnO & NiO &\\
 \hline
   & & & \\
 $J^{\uparrow\uparrow}_1$ & $-$5.8 & 0 \\
 $J^{\uparrow\uparrow}_4$ & $-$0.2 &$-$0.2\\
 $J^{\uparrow\uparrow}_5$ & $-$0.2 & $-$0.2 \\
   \hline
 & & & \\
 $J^{\uparrow\downarrow}_1$ & $-$5.7 &  0.1\\
 $J^{\uparrow\downarrow}_2$ & $-$6.0 & $-$14.3\\
  \hline
  & & & \\
 Collinear Calculations & & &\\
 \hline
   & & & \\
 $J_1$ & $-$4.2& 1.2 \\
 $J_2$ & $-$4.4 & $-$14.0 \\
  \hline
  & & & \\
 Experiment\\
   \hline
  & & & \\
  $J_1$ & $-$5.28 \cite{Pepy:1974uq}& $-$1.38 \cite{Hutchings:1972ij}, 1.38 \cite{Shanker:1973hc}  \\
 $J_2$ & $-$5.58 \cite{Pepy:1974uq}& $-$17.32 \cite{Hutchings:1972ij}, $-$18.30 \cite{Shanker:1973hc} \\
  \hline
  \hline
  \end{tabular}
\end{center}
\label{table1}
\end{table} 

However the LSF gives the freedom to explore other distributions of spin-spirals that may reduce the number required to achieve convergence of the exchange parameters. We noticed that several of the ${\bf q}$-points in our generated Monkhorst-Pack grids were high symmetry points and moreover several points in the set shared the same symmetry. In order to get rid of any similarities between the ${\bf q}$-points in the sets we employed a pseudo-random number generator to generate sets of random ${\bf q}$-points, which were used to extract the exchange parameters. We found that only 7 ${\bf q}$-points in such a random distribution were required to reach the desired convergence. Now, of course using a random number generator we can in principal generate a suboptimal mesh such as the equidistant one even though this is an unlikely event. Fortunately by calculating the condition number of the least square problem before doing the actual total energy calculation, we can discard any such set at minimal cost. In practice the generated random ${\bf q}$-points give the same results limited by the accuracy of the ab-initio total energy calculations. 

Structural changes of magnetic origin are driven by the distance and angle dependence of the exchange parameters in the Heisenberg model. With the trigonal distortion the degeneracy of $J^{\uparrow\uparrow}_1$ and $J^{\uparrow\downarrow}_1$ is lifted since the distance to the neighbors within the [111] plane is larger than the neighbors outside the [111] plane. Indeed it has been assumed that changes in the nearest neighbor exchange parameters are the main reasons for the exchange-striction effect in the transition-metal monoxides~\cite{Svane:1990qf}. In order to study the effects of the trigonal distortion on the exchange parameters and related properties we applied a volume conserving tensor T with a distortion parameter $\delta$ on the lattice vectors.

\begin{equation}
T = \dfrac{1}{(1+3\delta)^{1/3}}\left(\begin{array}{ccc}(1+\delta) & \delta & \delta \\ \delta & (1+\delta) & \delta \\\delta & \delta & (1+\delta)\end{array}\right)
\end{equation}

\begin{table}[htb]
\caption{The exchange parameters larger than 0.1 meV for MnO calculated with the experimental trigonal structure [$\delta=-0.005$]. The exchange parameters are given in meV. The parameters U and J were set to respectively 6.9 eV and 0.86 eV .}
\begin{center}
\begin{tabular}{ccccc} 
  \hline
  \hline 
  \\
  \multicolumn{5}{c}{MnO exchange parameters (meV)} \\
  \\
  \hline
 \multicolumn{2}{c}{Theory} & & \multicolumn{2}{c}{Experiment} \\ 
\\
\hline
\\
 $J^{\uparrow\uparrow}_1$ & $-$5.2 & & $J^{\uparrow\uparrow}_1$ & $-$4.05 \cite{JPSJ.36.11} \\
 $J^{\uparrow\uparrow}_4$ & $-$0.2 & & --& --\\
 $J^{\uparrow\uparrow}_5$ & $-$0.3 & & --& --\\
 \\
   \hline
\\
 $J^{\uparrow\downarrow}_1$ & $-$6.2 & & $J^{\uparrow\downarrow}_1$ & $-$5.35 \cite{JPSJ.36.11}\\
 $J^{\uparrow\downarrow}_2$ & $-$5.9 & & $J^{\uparrow\downarrow}_2$ & $-$5.25 \cite{JPSJ.36.11} \\
 \\
  \hline
 \hline
 \end{tabular}
\end{center}
\label{mno}
\end{table}

\begin{figure}[h]
\begin{center}
\includegraphics[width=1.0\columnwidth]{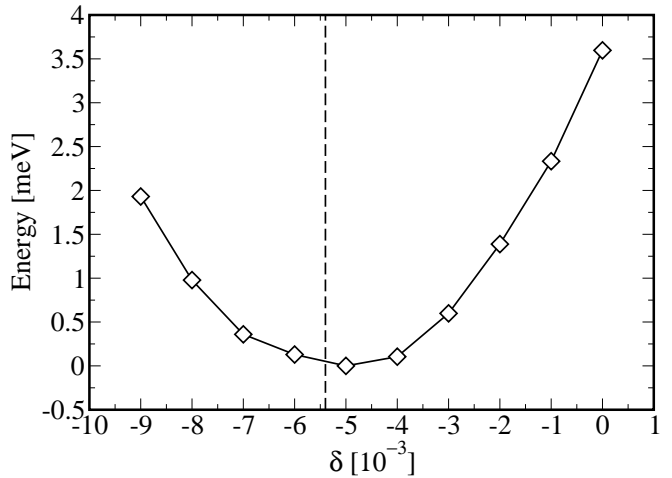}
\caption{The total energy as a function of the distortion parameter $\beta$ in MnO. The dashed line indicates the experimental trigonal distortion.\cite{Franchini:2005cr} The parameters U and J were set to respectively 6.9 eV and 0.86 eV.}
\label{strain}
\end{center}
\end{figure}

\begin{figure}[h]
\begin{center}
\includegraphics[width=1.0\columnwidth]{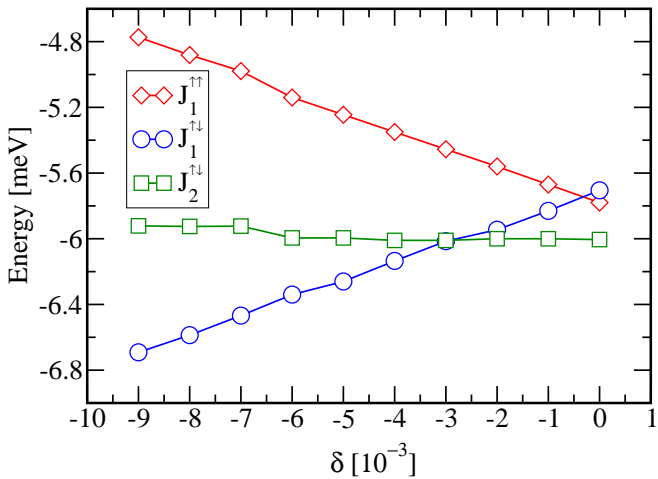}
\caption{(Color online)  The nearest neighbor exchange parameters as a function of the trigonal distortion in MnO. The parameters U and J were set to 6.9 eV and 0.86 eV respectively.}
\label{jdist}
\end{center}
\end{figure}

The changes in the exchange parameters that result in the trigonal ground state are challenging to calculate in the case of NiO as the differences in total energies are less than 0.1 meV. This result was also expected since the nearest neighbor exchange parameter is $\sim$0 meV. MnO on the other hand is a relatively easier case with an energy difference of 3.5 meV between the cubic structure and the trigonal ground state as shown in Fig.~\ref{strain}. Indeed the exchange parameters of MnO show a significant dependence on the distortion parameter as seen in Fig.~\ref{jdist}. The exchange parameters of trigonal MnO are summarized together with experimental results in table II. It is interesting to note in Fig.~\ref{TmOC} and Fig.~\ref{jdist} that neglecting exchange parameters beyond the second nearest neighbor introduces an error of the same order of magnitude as the changes introduced by the structural phase transition. 
\\

\begin{figure}
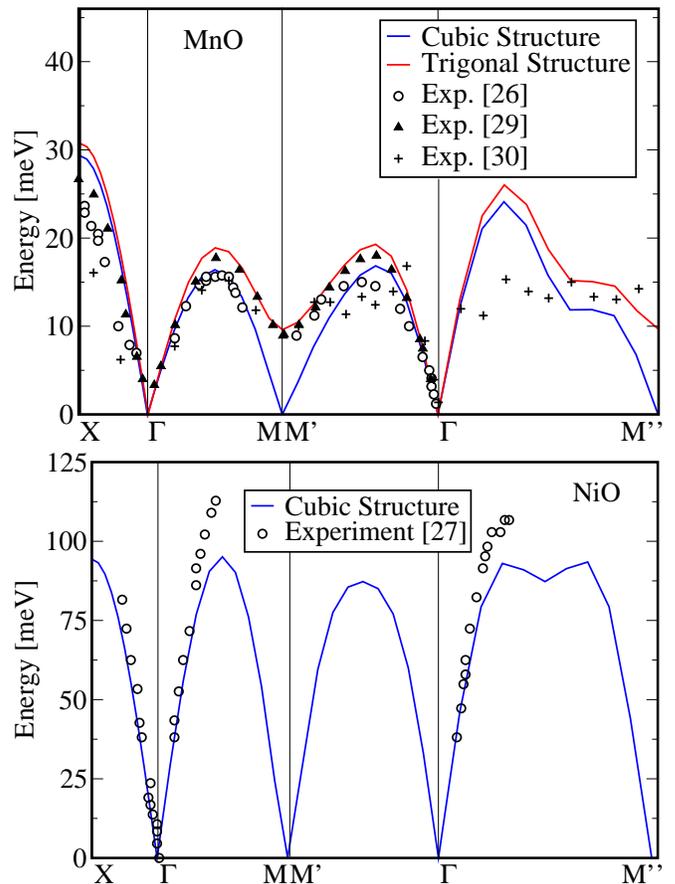

\begin{tabular}{c}
\includegraphics*[width=1.0\columnwidth]{MnOmagnon.eps}\\
\includegraphics*[width=1.0\columnwidth]{NiOmagnons.eps}
\end{tabular}
\caption{(Color online) Magnon dispersion curves for MnO and NiO. In the first case, we consider both the ideal rock salt structure and a trigonal structure defined by a distortion parameter  $\delta =-0.05$. The parameters U and J were set to respectively 6.9 eV and 0.86 eV  for MnO and 8.0 eV and 0.95 eV for NiO. }
\label{exp-magnon}
\end{figure}

\begin{figure}
\begin{tabular}{c}
\includegraphics*[width=1.0\columnwidth]{MnOmagnonC.eps}\\
\includegraphics*[width=1.0\columnwidth,bb = 2 40 425 323,clip=true]{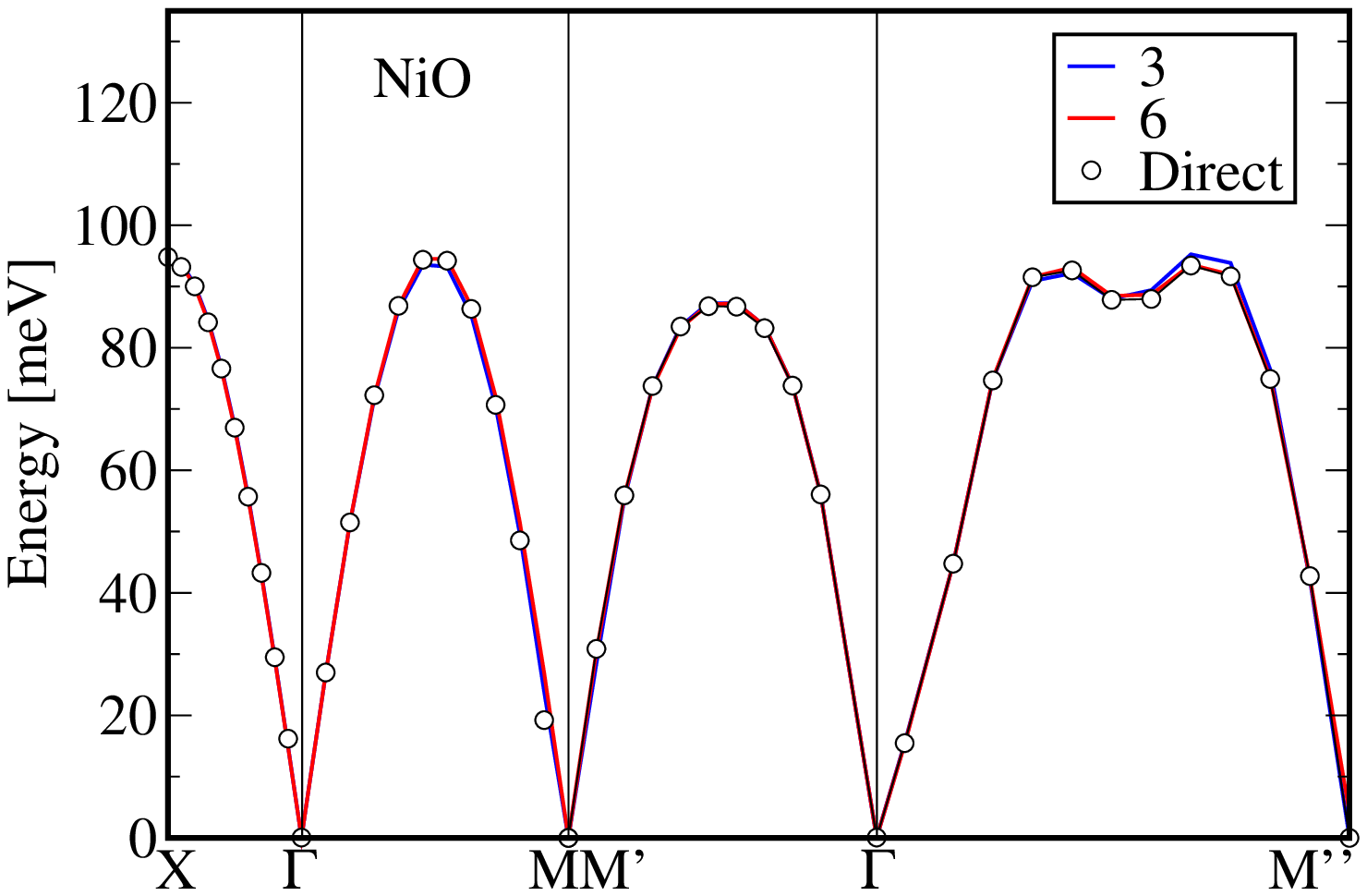}
\end{tabular}
\caption{(Color online) Magnon dispersion curves for MnO and NiO in the cubic structure for different numbers of included exchange parameters using Eq~\eqref{eqn8}. We also included the curve calculated with Eq.~\eqref{emagnon1} and Eq.~\eqref{emagnon2} in order to show that the magnon dispersion curves seems to be fully converged. The parameters U and J were set to respectively 6.9 eV and 0.86 eV for MnO and 8.0 eV and 0.95 eV for NiO. }
\label{conv-magnon}
\end{figure}

\begin{figure}[h]
\begin{center}
\includegraphics[width=0.50\textwidth]{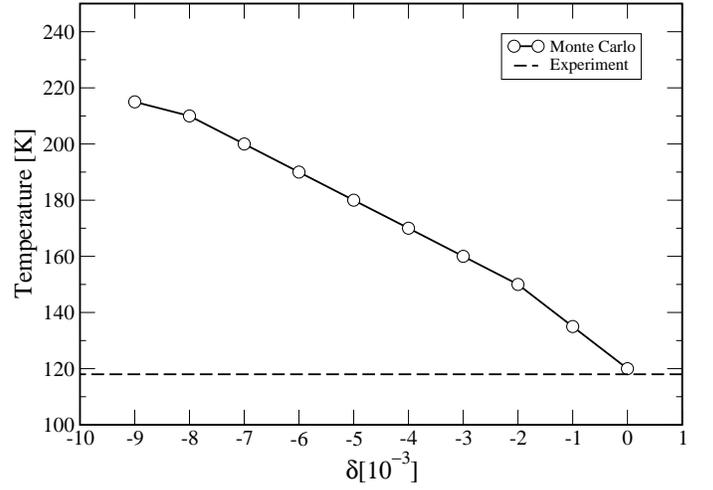}
\caption{Critical temperatures vs. trigonal distortion in MnO calculated from Monte Carlo simulations. Exchange parameters were taken from LDA+U calculations with parameters U and J set to 6.9 eV and 0.86 eV respectively.}
\label{Tn}
\end{center}
\end{figure}

In Fig.~\ref{exp-magnon}, we show the magnon dispersion curves for NiO and MnO in the cubic structure and in addition we consider the curve for the experimental trigonal structure in the latter case. We include all exchange parameters that are of the order of 0.1 meV or larger.  The Cartesian coordinates of the high symmetry points given in units of $2\pi/a$ where a is the lattice constant are $X=[0.25,0.25,0.25]$,  $\Gamma=[0,0,0]$, $M=[-0.5,-0.5,0.5]$, $M'=[0,0,1]$ and $M''=[1,1,0]$. A fair agreement with the experimental results is obtained and somewhat improved when the experimental structure is assumed for MnO.  

We can show that the non-zero energy obtained experimentally\cite{Pepy:1974uq, JPSJ.36.11, Goodwin:2007kx} at the $M/M'$-point in MnO when considering the trigonal structure is due to the changes in the exchange parameters introduced by exchange-striction.  In Fig.~\ref{conv-magnon} we show the convergence of the magnon energies as a function of included exchange parameters and compare these with results produced by using Eq.  \eqref{emagnon1} and \eqref{emagnon2}  that represent the infinite limit. In both cases, minor changes can be seen when increasing the number of exchange parameters from 3 to 6 for both materials but further increases will not introduce noticeable changes in the dispersion curves.

Finally, we present the results of Monte Carlo simulations. These calculations were done for both materials using a 9x9x9 supercell. A critical temperature of 420 K was obtained for NiO which is lower than the experimental value of 523 K. This result is expected since the magnitude of the second nearest neighbor exchange interaction is somewhat underestimated according to Table.~\ref{table1}. Our result is thus similar to the one calculated with a self-interaction correction scheme\cite{Fischer:2009bs} (458 K) since that functional also underestimates the magnitude of the second nearest neighbor exchange interaction. 

\begin{figure}[h]
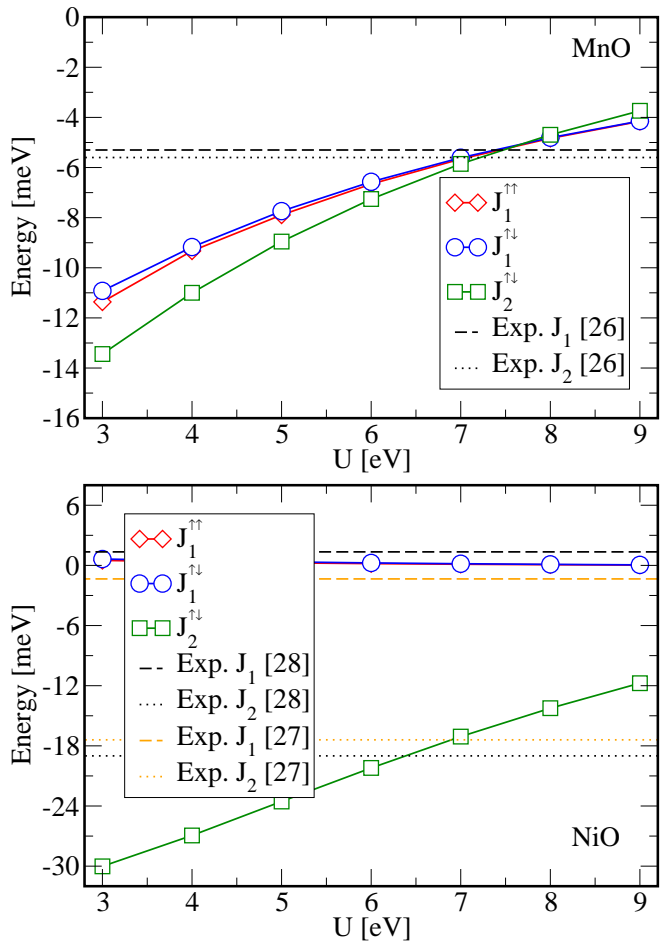

\centering
\begin{tabular}{c}
\includegraphics*[width=1.0\columnwidth]{MnOU.eps} \\
\includegraphics*[width=1.0\columnwidth]{NiOU.eps} \\
\end{tabular}
\caption{(Color online) The nearest and second nearest neighbour exchange parameters in MnO and NiO for different values of the Hubbard U parameter. Experimental values are given with dotted lines for nearest neighbor and dashed lines for next nearest neighbor exchange parameters. The two curves for the exchange parameters $J^{\uparrow \uparrow}_{1}$ and $J^{\uparrow \uparrow}_{2}$ are practically on top of each other.} 
\label{JUdep}
\end{figure}

In Fig.~\ref{Tn}, we show the calculated critical temperatures as a function of the distortion parameter $\delta$ for MnO. The experimental transition temperature is well reproduced for the ideal rock salt structure but as the distortion parameter  $\delta$ is increased the transition temperature rises which suggests that the application of pressure along the [111] direction will stabilize the magnetic ordering. This result is expected since the exchange parameters and excitation energies are well described as seen in Table.~\ref{table1} and Fig. \ref{exp-magnon}. 

In Fig.~\ref{JUdep} we show our results for the nearest and second nearest neighbor exchange parameters for values of Hubbard U from 3~eV to 9~eV. For MnO it is again seen that U $\simeq$ 7 eV reproduces the experimental situation for the LDA+U functional while for NiO it is clear that experimental $J_2$ is obtained for the range of U $\simeq$ 6 eV - 7 eV. We may thus expect that also a closer fit of the calculated dispersion curves and critical temperatures are obtained with such a lower value of Hubbard U. 
\\

It can be seen that the curves of $J^{\uparrow\uparrow}_1$ and $J^{\uparrow\downarrow}_1$ diverges up to 0.5 meV from each other for low values of U which indicate that those calculations are not fully converged with respect to the number of k-points. In principle one should converge the calculations with respect to all relevant parameters for each value of U, but we deemed such a thorough convergence needlessly time consuming since we were primarily interested in the range close to the experimental values and the overall trend. 

\section{Discussion}
In this study we have derived a set of equations to extract the exchange parameters of the Heisenberg model from spin-spiral total energies using a LSF procedure. The results were compared with those of the previously implemented FT based method. 

For our studied transition metal monoxides, 11 ${\bf q}$-points placed in an equidistant mesh have to be considered  for the LSF while 29  ${\bf q}$-points were required for the FT method in order to obtain the exchange parameters. 

In comparison to the FT method we have more flexibility in the selection of ${\bf q}$-points with a LSF method that potentially can reduce computational time even further. But as in all LSF methods we have to ensure that the problem we set up is well conditioned in order to obtain sensible results. Fortunately the conditioning can be known prior to making any total energy calculation which makes the problem manageable. By considering sets of random ${\bf q}$-points we reduced the required amount of ${\bf q}$-points from 11 to 7 and hence computational time is reduced. This gain is probably related to the high symmetry of the points in the generated equidistant meshes and is more pronounced the more sparse the mesh is. 

Thus when we consider systems with short ranged interactions such as insulators or half-metals where the corresponding ${\bf q}$-points sets are relatively small compared to sets appropriate for metals we expect larger gains by using such randomised sets. 

While the calculations of exchange parameters from a set of collinear magnetic structures can give reasonable results for some cases, e.g., NiO, other systems like MnO shows less robust properties and errors of the same scale as the exchange parameters are introduced. 

Using our calculated exchange parameters, we extract magnon dispersion curves and critical temperatures and find a good correspondence to experiments. For NiO the value of U used in the LDA+U functional seems to be slightly overestimated resulting in slightly lower values of maxima in magnon dispersion curves compared to experiments and an underestimated critical temperature. In the case of MnO, the chosen values of  6.9 eV for U and 0.86 eV for J are shown to reproduce experimental results with respect to both magnetic properties and the related structural distortion. However, we have observed that the choice of the ideal rock salt structure or the experimental trigonal structure for the calculations of exchange parameters is crucial for the resulting critical temperatures and magnon dispersion curves. 

Exchange-striction in the form of a contraction along the [111] direction stabilizes the magnetic ordering of the  domain with antiferromagnetic ordering in the [111]  direction which we can see in the increasing critical temperature calculated with the Monte Carlo method. The destabilization of the previously equivalent domain with ordering in the  $[\bar{1}\bar{1}1]$ direction can be seen in the changes in the magnon dispersion curves. In the ideal cubic structure the equivalence of the directions is shown by the degeneracy of the magnon energy at the $\Gamma$ and the $M/M'$ -point. With the introduction of the trigonal distortion the magnon energy at the the $M/M'$ -point becomes non-zero and the degeneracy is lifted. Applied to the transition-metal monoxides we show that the magnon energies are fully converged with the inclusion of 6 exchange parameters. 

We investigate how well the LDA+U functional can describe magnetic properties in these two materials by a variation of the Hubbard U parameter and conclude that the LDA+U functional can reproduce experimental exchange parameters with a suitable choice of U. For NiO, the Hubbard U required to reproduce these experimental results is in the same range as the U-parameter used to reproduce experimentally obtained electron loss spectra, structural parameters and optical properties ~\cite{Dudarev:1998ys,Bengone:2000zr}. We thus find that the lower Hubbard U's as estimated by linear response~\cite{Cococcioni:2005fk} and constrained RPA~\cite{Karlsson:2010kx} seems to give a reasonable description of the system for a range of measured properties. For the case of MnO there are less theoretical results available that would single out a particular value of Hubbard U. But it seems that in this case constrained LDA calculations yield a value of Hubbard U that gives a good description of magnetic and structural properties. It is thus not clear if the preference to a particular method of obtaining Hubbard U give systematically better agreement with experiments.

Furthermore that a single value of Hubbard U exists that results in a satisfactory description of all available experimental properties is not always the case~\cite{Schlipf:2011uq}. This points to the limitation of the LDA+U method itself. While the method is roughly as fast as standard DFT-calculations employing the LDA or GGA potentials, other more advanced and computationally more expensive methods such as DMFT or self consistent GW calculations are expected to give a more accurate description of the electronic structure. ~\cite{Kunes:2007oq, Faleev:2004vn}. It would be interesting to combine the presented mathematical framework  with other more recently developed electronic structure methods that provide accurate descriptions of strongly correlated systems. 

 We generalized the well known relationship between spin-spiral total energy differences and magnon energies for systems containing a single magnetic sublattice to the case of multiple magnetic sub lattices. This direct way of calculating magnon energies from spin-spiral total energy differences might be useful if the magnon energies at a specific ${\bf q}$-point is needed since the number of spin-spiral total energies to calculate the magnon energies at a specific ${\bf q}$-point is greatly reduced compared to the approach where all exchange parameters must be calculated. The advantage grows with the number of sizeable exchange parameters. If the specific ${\bf q}$-point is a high symmetry point then symmetry operations compatible with the ${\bf q}$-point might be employed for the {\em ab initio} calculations reducing computational time even further. 

\section{Acknowledgments}

\begin{acknowledgments}
This work was supported by the Young Investigators Group Programme of the Helmholtz Association, Germany, contract VH-NG-409. 
We gratefuly acknowledge the support of J\"{u}lich Supercomputing Centre.
\end{acknowledgments}

\bibliographystyle{apsrev4-1}
\bibliography{exch-par}

\end{document}